\documentclass[conference]{IEEEtran}
\IEEEoverridecommandlockouts
\usepackage{cite}
\usepackage{amsmath,amssymb,amsfonts}
\usepackage{algorithmic}
\usepackage{graphicx}
\usepackage{textcomp}
\usepackage{xcolor}
\def\BibTeX{{\rm B\kern-.05em{\sc i\kern-.025em b}\kern-.08em
    T\kern-.1667em\lower.7ex\hbox{E}\kern-.125emX}}

\usepackage{makecell}
\usepackage{amsmath}
\usepackage{amsmath, amssymb, amscd, amsthm, amsfonts}
\usepackage{colortbl}
\usepackage{soul}
\usepackage{multicol}
\usepackage{comment}
\usepackage[colorlinks=true,linkcolor=blue,anchorcolor=black,citecolor=red,filecolor=black,menucolor=black,runcolor=black,urlcolor=black]{hyperref}
    
\begin{document}

%\title{Large Language Models for Cybersecurity of Digital Substations in Smart Grids\\

\title{ChatGPT and Other Large Language Models for Cybersecurity of Smart Grid Applications\\

%\thanks{This work was supported by the Korea Institute of Energy Technology Evaluation and Planning (KETEP) and the Ministry of Trade, Industry and Energy (MOTIE) of the Republic of Korea under grant No. 20206910100020.}

%\thanks{}
%\thanks{Corresponding author: Yong-Hwa Kim (e-mail: yongkim@ut.ac.kr).}

\thanks{A. Zaboli and J. Hong are with the Department of Electrical and Computer Engineering, University of Michigan -- Dearborn, Dearborn, MI 48128, USA. 

S. L. Choi is with the Power Systems Engineering Center, National Renewable Energy Laboratory (NREL), Golden, CO 80401, USA.

T.-J. Song is with the Department of Urban Engineering, Chungbuk National University, Cheongju 28644, South Korea.}
}

\author{\IEEEauthorblockN{Aydin Zaboli, 
\textit{Graduate Student Member, IEEE}, 
Seong Lok Choi, 
\textit{Member, IEEE}, 
Tai-Jin Song, 
\textit{Member, IEEE}, \\
Junho Hong, 
\textit{Senior Member, IEEE}
}}

\maketitle

\begin{abstract}
Cybersecurity breaches targeting electrical substations constitute a significant threat to the integrity of the power grid, necessitating comprehensive defense and mitigation strategies. Any anomaly in information and communication technology (ICT) should be detected for secure communications between devices in digital substations. This paper proposes large language models (LLMs), e.g., ChatGPT, for the cybersecurity of IEC 61850-based communications. Multi-cast messages such as generic object oriented system events (GOOSE) and sampled values (SV) are used for case studies. The proposed LLM-based cybersecurity framework includes, for the first time, data pre-processing of communication systems and human-in-the-loop (HITL) training (considering the cybersecurity guidelines recommended by humans). The results show a comparative analysis of detected anomaly data carried out based on the performance evaluation metrics for different LLMs. A hardware-in-the-loop (HIL) testbed is used to generate and extract dataset of IEC 61850 communications.

\end{abstract}


\section{Introduction}
Digital substations serve as crucial elements within modern power systems, characterized by their escalating complexity and integration. 
%The incorporation of advanced technologies and protocols, notably GOOSE and SV, has emphasized the urgent need to enhance cybersecurity defenses. 
%These protocols are fundamental to ensuring the instantaneous responsiveness and dependable functionality of digital substations. However, these are vulnerable to an array of cyber threats, including the intrusion of anomalous data, which poses a significant risk to the structural integrity and operational efficacy of the system~\cite{hong2022automated}. 
GOOSE and SV are instrumental in facilitating rapid and dependable communication in the context of digital substations. Nevertheless, the open architecture intrinsic to these protocols makes them vulnerable to cyberattacks. 
%There has been significant growth in scholarly work focused on detecting and mitigating anomalous activities within GOOSE and SV datasets to detect or mitigate cyber threats. 
The focal point of scholarly endeavors is the refinement and implementation of complex algorithms tailored for the contemporaneous oversight and scrutiny of network traffic~\cite{hussain2023novel}.
%\cite{reda2021vulnerability}
%This is aimed at discerning and mitigating irregular patterns that are symptomatic of impending cyberattacks 
%Substations are at a risk of cyberattacks due to their use of Ethernet communications for data transfer. Hackers can exploit this vulnerability to disrupt operations. From January through August 2022, nearly 101 cyberattacks occurred across the U.S. on electrical equipment. It is crucial for facilities, vital to the power supply, to have a strong security to protect against such threats~\cite{darkreading}.

%Cyberattacks, especially coordinated ones, can lead to significant power outages. Therefore, implementing comprehensive security measures, including layered defense strategies, tracking attack progressions, and using intelligence-driven approaches, is essential to protect these systems
Intrusion detection system (IDS)-based machine learning (ML) methods have been the foundation for detecting and mitigating anomalies in GOOSE and SV messages. While these methods offer precision and are data-driven, they come with a significant challenge. Every time a new attack pattern emerges, the models need to be re-trained. This necessity for re-training consumes time and resources and leaves the system vulnerable during the interim periods when the new threats are not yet incorporated into the model's knowledge base~\cite{hong2022automated}.
%, mujeeb2021machine}
On the other hand, LLMs such as ChatGPT 4.0 offer a more dynamic and adaptable approach. Unlike ML models, LLMs are designed to understand context, allowing them to recognize and respond to novel threats even if they have not been explicitly trained in them. This contextual understanding minimizes the efforts required in the face of evolving cyber threats. Instead of frequent re-training sessions, LLMs can interpret and adapt to new information, providing a more resilient and efficient solution for anomaly detection in digital substations~\cite{gill2023chatgpt, ten2011anomaly}.
%% After literature survey %%%%%%%%%
In the area of cybersecurity for digital substations, LLMs can play a pivotal role in anomaly detection, enhancing the security layers. These models can analyze vast datasets, identify patterns, and detect anomalies indicative of potential cyberattacks. These models are designed to investigate through extensive data, including GOOSE and SV messages, to effectively distinguish regular patterns from irregularities~\cite{gupta2023chatgpt}. The incorporation of artificial intelligence (AI) aids into real-time monitoring is crucial to accelerate responses to security breaches.
%ML models can enhance intrusion detection in digital substations by analyzing patterns and anomalies in vast datasets. 
%This allows for real-time detection of cyberattacks, ensuring grid stability and security. 
%In~\cite{alvee2021ransomware}, the authors proposed an AI-based ransomware detection approach that utilizes a convolutional neural network. The unique aspect of this approach is the conversion of binary files into 2-D image files for detection. Experimental results indicate a high detection accuracy of 96.22\%. 
A unique deep learning-based system tailored for detecting cyberattacks on protective relays was developed based on extensive real-world datasets~\cite{khaw2020deep}. GOOSE and SV messages are vulnerable to replay and message injection attacks, involving the re-transmission of unaltered messages or the transmission of fake, malicious ones. These attacks disrupt system operations either by replaying old messages or by injecting new, deceptive messages that mimic legitimate behavior~\cite{10339874, 10256104}. However, the diversity and complexity of cyberattacks necessitate advanced detection mechanisms. Also, balancing the model's sensitivity to detect minor anomalies while avoiding false positives (FPs) is crucial.
%An advanced ML technique for real-time anomaly detection was proposed in~\cite{panthi2020anomaly}. It employed feature extraction techniques to optimize the data; however, handling the vast amount of data without compromising the system's performance is challenging. 
%Moreover, a trade-off between the detection accuracy, computational efficiency and robustness against evolving cyberattack patterns is controversial. 
In~\cite{jay2022unsupervised}, a novel unsupervised learning approach for an IDS of GOOSE messages is suggested based on a combination of autoencoders and clustering techniques for efficient detection. 
%A combination of CNN and long short-term memory networks proposed for an enhanced performance in~\cite{ankitdeshpandey2020development}. %In~\cite{upadhyay2020gradient}, authors introduced a gradient boosting method for the feature selection in intrusion detection system which it demonstrates improved accuracy and reduced FPs in detecting anomalies. A novel IDS designed to combat Manufacturing Message Specification (MMS)-based measurement attacks which integrated advanced detection algorithms to improve the accuracy~\cite{zhu2020intrusion}. Authors in~\cite{ustun2021machine} suggested a novel IDS which employs a combination of convolutional neural networks (CNN) and long short-term memory (LSTM) networks to capture both spatial and temporal patterns in data.
According to literature surveys, challenges in the applicability of ML models in IDSs can include ensuring the reliability and robustness of the model in real-time power grids considering new cyberattacks, a trade-off between complexity and accuracy due to large datasets, and the adaptability of the ML model to evolving cyberattacks and changing the substation infrastructure. Furthermore, a re-training process is required for new cyberattacks; however, LLMs can handle these challenges effectively and reduce the processing time.

%\subsection{Contributions}
This paper proposes for the first time the employment of LLMs based on HITL interactions to detect anomalies in GOOSE and SV datasets for cybersecurity considerations in substations. Hence, this paper focuses on the cybersecurity of multicast messages, and we will focus on other protocols in substations in the future. This paper suggests human recommendations for data pre-processing for these communication protocols. This process minimizes efforts (unlike applying ML methods) when encountering new cyberattacks (or anomalies). It does not affect the model's complexity/precision and is faster to implement. Moreover, this paper makes a comparison between LLMs (i.e., ChatGPT 4.0~\cite{chatgpt4.0}, Anthropic's Claude 2~\cite{AnthropicClaude2}, and Google Bard/PaLM 2~\cite{googlebard}) to evaluate their performance. The actual datasets for GOOSE and SV packets are extracted from the HIL testbed. The main contributions of this paper can be summarized as follows:
\begin{itemize}
    \item This paper proposes the usage of different LLMs in the cybersecurity of digital substations in terms of performance evaluation metrics.
    \item LLM-based HITL is considered an IDS to detect abnormal data in IEC 61850 communication protocols.
    \item A conversion of the IDS algorithm to text is employed for training datasets to detect anomalies in LLMs.
\end{itemize}
%\subsection{Paper Structure}
The remainder of the paper is organized as follows: Section II states a representation of the cybersecurity of digital substations using LLMs. The proposed HITL technique, along with the feature extraction and analysis of datasets, are mentioned in Section III. Section IV presents the results and discussion of the evaluation metrics according to different levels of training. Finally, this paper is deduced in Section V.

\section{Cybersecurity of Digital Substations Using Large Language Models}
%As digital substations integrate advanced communication technologies, they become potential targets for cyberattacks. 
%Ensuring the cybersecurity of critical infrastructure involves implementing multi-layered protective measures, ranging from physical access controls to advanced IDS. Regular assessments are essential to combat evolving cyber threats and secure the reliability and stability of power grids. 
%The following section presents the HIL testbed along with the LLM-based IDS.
\subsection{Cybersecurity of Digital Substations}
\begin{figure}[!t]
\centerline{\includegraphics[width=0.9\columnwidth]{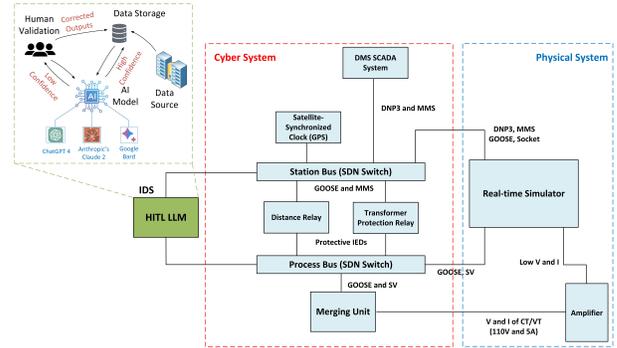}}
\caption{HIL Testbed considering the IDS with human recommendations.}
\label{fig:HIL TESTBED}
\end{figure}

A cyber-physical power system testbed serves as an instrumental platform for studying the causal relationships associated with cyber intrusions, the robustness of power systems, and the dependability of applications in a realistic environment. Within such a real-time HIL testbed, all constituent elements, encompassing hardware, software, communication mechanisms, and emulators, are coordinated in alignment with the global positioning system (GPS). The real-time dynamics pertinent to communication and information processing become imperative in the context of analyzing cyber intrusions, detection mechanisms, and mitigation strategies~\cite{hong2015cyber}. As seen in Fig.~\ref{fig:HIL TESTBED}, the testbed consists of protective intelligent electronic devices (IEDs), software-defined networking (SDN) switches, a satellite-synchronized clock, a merging unit, a supervisory control and data acquisition (SCADA) system, a real-time digital simulator, and the amplifier.
%Several products exist that generate both analog and digital outputs. These products are extensively employed for hardware diagnostic evaluations and the simulation of power systems. 
The distributed management system (DMS) SCADA system can get measurements and issue a control command via DNP3 communication.
%~\cite{ABB_MicroSCADA}. 
Various IEDs are implemented, including the merging unit IED and protective IEDs. These IEDs possess the proficiency to transmit control commands (e.g., GOOSE messages) pertinent to a circuit breaker (CB). Conversely, a CB IED (modeled in a real-time simulator) is specifically configured to subscribe to GOOSE messages and publish the status (open or closed) to protective IEDs. The merging unit IED has the ability to forward digital current and voltage values (i.e., SV), taking into account the amplifier from the digital simulator, to the protective IED. Furthermore, the proposed HITL LLM-based IDS is engineered to identify anomalies and potential security threats within the substation automation framework and maintains a connection to SDN switches~\cite{hong2015cyber}.
The purpose of this paper is to demonstrate an IDS considering the LLM-based HITL process. Hence, the GOOSE and SV packets are extracted from the HIL testbed for further analysis in different LLMs considering the human recommendations that are described in the subsequent section. 
%This study is not intended to go into detail on the HIL testbed's internal processing.
\subsection{Large Language Model-Based Human-in-the-loop Process}
Generative AI (GenAI) models, constructed through deep neural network methods, are designed to discern patterns and structures from extensive training datasets, subsequently producing similar content. The capabilities of GenAI encompass the creation of diverse content types, ranging from text and images to sounds and various data forms. The introduction of ChatGPT has markedly influenced the broader AI/ML domain, exemplifying GenAI's potential to resonate with the wider populace and altering prevailing conceptions of AI/ML. The technological sector is actively pursuing the refinement of advanced LLMs aimed at simulating authentic human interactions, as evidenced by innovations (e.g., Microsoft's GPT and Google Bard/PaLM 2). Over the past year, GenAI has strengthened its presence as a prevalent online tool~\cite{gupta2023chatgpt}. 

LLMs and GenAI systems present considerable opportunities to augment productivity and operational efficiency. However, their application, especially in sectors characterized by high risk and stringent regulations, brings about notable challenges. A potential strategy to mitigate risks is adopting the HITL process, as illustrated in Fig.~\ref{fig:HIL TESTBED} by the HITL LLM box. Incorporating human interactions during training, validation, and testing stages can expedite the learning process and improve the confidence level of outputs. Initially, individuals can explain the execution of specific tasks and subsequently offer insights into the model's efficacy. This involvement can be manifested in modifying the model’s results. Drawing insights from a fusion of human demonstrations and assessments has proven to surpass the efficiency and speed of ML methods. The HITL paradigm becomes indispensable when confronted with constraints (e.g., when data presents anomalies or lacks comprehensiveness), leading to uncertainties about the model's capability to address all scenarios. Moreover, consistent human oversight and verification are useful, especially when inaccuracies in model predictions could have severe consequences~\cite{HITL}. In the proposed model, there are human recommendations to improve the model efficiency based on GOOSE and SV message features. Thus, this method is helpful in minimizing the trials by entering new data into the normal dataset and avoiding the re-training process. Also, the adaptability and robustness of models can be improved quickly.
\begin{comment}
\begin{figure}[!t]
\centerline{\includegraphics[width=0.9\columnwidth]{images/HITL LLM.png}}
\caption{An LLM-based HITL as an IDS in cybersecurity of digital substations.}
\label{fig:HITL}
\end{figure}
\end{comment}
%\subsection{Challenges of LLMs in Cybersecurity Considerations}

Allowing a language model unrestricted access to data pertaining to critical infrastructure necessitates meticulous study, given the significant security and privacy implications. Implementing accurate access controls and encryption and authentication protocols is imperative to mitigate unauthorized data access. 
%Notably, LLMs exhibit certain constraints. Their capability to discern intricate nuances within cybersecurity or to provide precise responses in particular scenarios might be limited. 
It is vital for human specialists to exercise continuous supervision and assess the outputs of the model, ensuring the validity and dependability of AI-facilitated cybersecurity methodologies.
In addition, there exists the potential for LLMs to unintentionally disclose confidential information during engagements, especially if they lack appropriate training or protection~\cite{gupta2023chatgpt}.
%Attackers might also attempt to leverage the natural language processing abilities of LLMs to gain insights on executing systemic disruptions by manipulating specific power system apparatuses. Such endeavors become markedly risky if the model elucidates comprehensive methodologies for malicious intentions. 
The cybersecurity of LLMs is out of scope for this research, and the purpose is solely to employ LLMs as tools for detecting anomalies in communication messages.

\section{IEC 61850-based Communication Datasets and Human Recommendations Process}

\subsection{GOOSE and SV Datasets}
\begin{figure}[!t]
\centerline{\includegraphics[width=0.7\columnwidth]{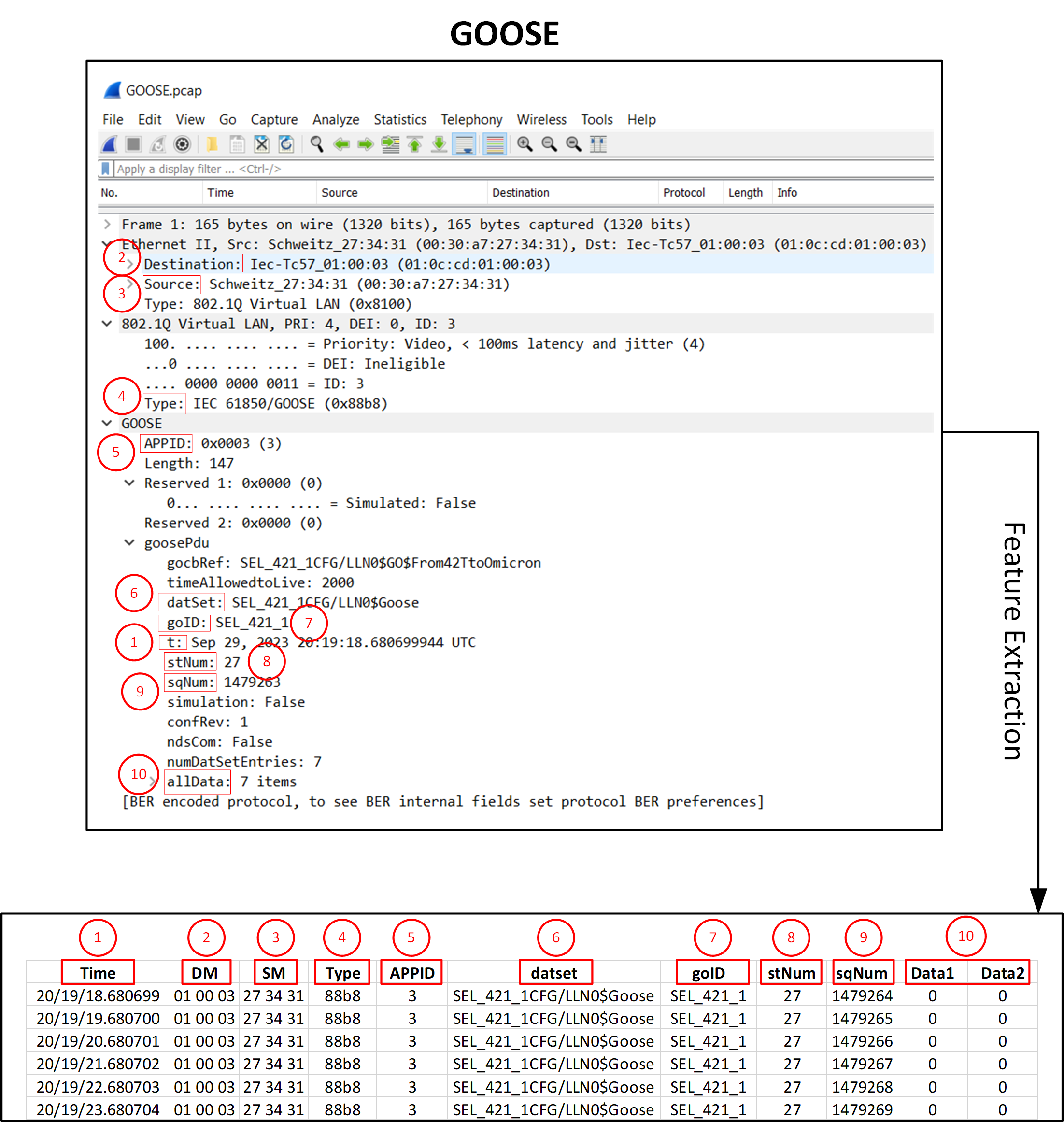}}
\caption{A pre-processing step based on the feature extraction for a log of GOOSE message (actual data from an HIL testbed).}
\label{fig:Pre-processing}
\end{figure}

The GOOSE and SV packets are extracted from the HIL testbed. The ``.pcap'' means that these packets are captured using Wireshark (a network packet analyzer), as seen in Fig.~\ref{fig:Pre-processing}. 
As shown, there are 10 data points for GOOSE packets based on the extracted features. The SV datasets follow the same procedures, with the 7 most important features as dataset columns. ``Time'' shows the time at which the packet is actually sent, and the format of this feature is based on hour, minute, and second (including microsecond level). The features ``DM'' and ``SM'' refer to the destination and source media access control (MAC) addresses, respectively. This specific ``DM'' address (01 00 03) of GOOSE messages shows the target devices (sent to the device that subscribes to this MAC address). Also, the ``SM'' address of GOOSE messages is 27 34 31 which shows the sender's IED. The indicators for GOOSE and SV are shown as ``type,'' which is 88 b8 and 88 ba, respectively. The ``APPID'' values for GOOSE and SV communications are 3 and 40, respectively. Also, ``datSet'' and ``goID'' are assigned based on ``DM'' and indicate the dataset name and GOOSE identification, respectively. Based on Fig.~\ref{fig:Pre-processing}, there is a GOOSE block reference (``gocbRef'') that indicates the name of the GOOSE in the ``goosePdu.'' ``stNum'' and ``sqNum'' express the state and sequence numbers in GOOSE communications, respectively. Furthermore, two data types (``data1'' and ``data2'') are considered based on GOOSE packets. In the SV dataset, there are ``svID'' and ``smpCnt'' in the ``savPdu,'' which indicate SV identification and sample count number. A large number of datasets have been used to train the GOOSE and SV communications to check the performance evaluation of LLMs.
\subsection{Human Recommendations for Intrusion Detection Systems}
According to the given datasets, a series of human recommendations based on the violations in the GOOSE and SV datasets can be described.
Some different attacks and errors were considered, such as the data injection (DI) attack, the denial-of-service (DoS) attack, the system problem, and the replay (RE) attack for GOOSE and SV communications. These attacks can be described as follows. A failure to satisfy at least one recommendation leads to the relevant attack. Regarding the DoS attack for SV, the sample/cycle is $80$ and the frequency is $60$ Hz, so there are a total of $4,800$ samples per second. If $1/4800$ is calculated, $208$ microseconds can be achieved. Hence, the normal time for a DoS attack should be around this time. The process can be done for GOOSE messages as well. The ``heartbeat'' of GOOSE packets refers to a regular, periodic message sent over the network to indicate the status of the system. The heartbeat (e.g., $2$ seconds) ensures continuous monitoring and quick detection of any changes or failures in the system. The frequency of these heartbeat messages can vary depending on the configuration and requirements of the specific substation system. Typically, GOOSE messages are sent at intervals ranging from a few milliseconds to several seconds. Regarding SV packets, a heartbeat indicates the operational status or health of a system. The exact frequency or interval of the heartbeat for SV packets can vary depending on the specific implementation and requirements of the digital substation system.
\begin{itemize}
    \item \textbf{Attacks/errors on GOOSE datasets}
    
    - \textbf{DI}: If data has the same ``DM'' and ``SM,'' ``sqNum'' should be increased every time.
    
    - \textbf{DI}: If there is any change in ``data1'' or ``data2,'' ``stNum'' should be increased by 1 and ``sqNum'' should be reset to 0.
    
    - \textbf{DI}: If data has the same ``DM'' and ``SM,'' once ``stNum'' is increased, it cannot go back to smaller numbers.
    
    - \textbf{DoS}: There are up to 10 packets (rows) within 10 ms.
    
    - \textbf{System Problem}: There should be a packet (dataset) within 10 s.
    
    - \textbf{RE}: If there is any change in ``data1'' or ``data2,'' ``stNum'' should be increased 1 and ``sqNum'' should be reset to 0.
\end{itemize}

\begin{itemize}
    \item \textbf{Attacks/errors on SV datasets}
    
    - \textbf{DI}: The range of ``smpCnt'' is from 0 to 4799.
    
    - \textbf{DI}: Once the ``smpCnt'' is increased, it should be increased up to 4799 and then reset to 0.
    
    - \textbf{DI}: ``smpCnt'' cannot be decreased until it reaches 4799 and resets to 0.
    
    - \textbf{DoS}: A normal time interval should be around 208 ms.
    
    - \textbf{DoS}: There are up to 12 packets within 2.083 ms.
    
    - \textbf{System Problem}: ``smpCnt'' should be increased every time by 1.
\end{itemize}
The recommended considerations are applied to datasets to train LLMs. This helps to improve the accuracy of the pre-trained model based on the ML model, even though there is new data. The purpose is to show the performance evaluation based on datasets generated at three different levels, including a dataset without training (without human recommendations), with partial training (recommendations of DI and DoS attacks), and with full training. Then, the performance evaluation metrics of different LLMs are compared. This process assists in minimizing the trials for re-training ML models and the adaptability of the model in cases where there is new data. The next section presents the performance evaluation results, considering the HITL process in different LLMs.
\section{Results and Discussion}
This section presents a comparison of results between LLMs at different levels. Precision is a preferable performance metric, denoting the rate at which an IDS correctly detects anomalies. However, relying solely on precision metric might be misleading when evaluating anomaly detection techniques, especially in scenarios characterized by significant differences between FPs and false negatives (FNs). Therefore, this section presents the performance analysis of different LLMs considering the HITL for the GOOSE and SV datasets. The fundamental performance metrics for anomaly detection analysis are described and discussed in this section. 
%\subsection{Performance Assessment}
\begin{comment}

Eqs.~(\ref{TPR})--(\ref{F1-score}) provide methodologies for fundamental performance metrics for anomaly detection analysis as follows~\cite{hong2015cyber}:
%\begin{equation} \label{Accuracy}
%    Accuracy = \frac{TP+TN}{TP+TN+FP+FN}
%\end{equation}
\begin{equation} \label{TPR}
    Detection \, Rate = TPR = Recall = \frac{TP}{TP+FN} 
    %\quad TNR = \frac{TN}{TN+FP}
\end{equation}
\begin{equation} \label{FPR/FNR}
    FPR = \frac{FP}{FP+TN}, \quad FNR = \frac{FN}{FN+TP}=1-TPR
\end{equation}
\begin{equation} \label{F1-score}
\small
    Precision = \frac{TP}{TP+FP}, \quad F1-Score = 2 \times \frac{Precision \times Recall}{Precision + Recall}
\end{equation}
\end{comment}
A description of evaluation metrics based on the detection of anomaly data in GOOSE and SV communication protocols, along with the results based on case studies, is shown in Table~\ref{results}. Due to the limitations of LLMs and computational speed, this paper focuses on online detection, not real-time intrusion detection. 
\begin{table*}[h]
\centering
%\fontsize{6pt}{8pt}\selectfont
\caption{A comparison of detection results (without, partial and full terms show the levels of training process).}
\label{results}
\begin{tabular}{|c|p{5.5cm}|ccccccccc|}
\hline
  \cellcolor[HTML]{656565}        & \textbf{\makecell{IEC 61850-based Communication}} & \multicolumn{9}{c|}{\cellcolor[HTML]{FFFFC7}{\color[HTML]{000000} \textbf{GOOSE}}}                                        \\ \hline
\cellcolor[HTML]{656565}      & \makecell{\textbf{LLMs}} & \multicolumn{3}{c|}{ChatGPT 4.0}                                                        & \multicolumn{3}{c|}{Anthropic's Claude 2}                                                        & \multicolumn{3}{c|}{Google Bard/PaLM 2}                                   \\ \hline
      \textbf{\makecell{Metrics}}    & \textbf{\makecell{Description}} & \multicolumn{1}{c|}{without} & \multicolumn{1}{c|}{partial} & \multicolumn{1}{c|}{full} & \multicolumn{1}{c|}{without} & \multicolumn{1}{c|}{partial} & \multicolumn{1}{c|}{full} & \multicolumn{1}{c|}{without} & \multicolumn{1}{c|}{partial} & full \\ \hline
\textit{TPR}       & \makecell{A ratio of correct GOOSE anomalies that were \\ correctly identified (also, named recall).} & \multicolumn{1}{c|}{78.18\%}        & \multicolumn{1}{c|}{85.45\%}        & \multicolumn{1}{c|}{98.18\%}     & \multicolumn{1}{c|}{78.18\%}        & \multicolumn{1}{c|}{83.64\%}        & \multicolumn{1}{c|}{89.09\%}     & \multicolumn{1}{c|}{74.5\%}        & \multicolumn{1}{c|}{81.8\%}        & 89.1\%  \\ \hline
%\textit{TNR}       & \makecell{A ratio of normal GOOSE data that were \\ correctly classified as clean data.} & \multicolumn{1}{c|}{52\%}        & \multicolumn{1}{c|}{68\%}        & \multicolumn{1}{c|}{96\%}     & \multicolumn{1}{c|}{44\%}        & \multicolumn{1}{c|}{56\%}        & \multicolumn{1}{c|}{68\%}     & \multicolumn{1}{c|}{44\%}        & \multicolumn{1}{c|}{60\%}        &   80\%   \\ \hline
\textit{FPR}       & \makecell{A ratio of normal GOOSE data that were \\ wrongly identified as anomalies.} & \multicolumn{1}{c|}{48\%}        & \multicolumn{1}{c|}{32\%}        & \multicolumn{1}{c|}{4\%}     & \multicolumn{1}{c|}{56\%}        & \multicolumn{1}{c|}{44\%}        & \multicolumn{1}{c|}{32\%}     & \multicolumn{1}{c|}{56\%}        & \multicolumn{1}{c|}{40\%}        &    20\%  \\ \hline
\textit{FNR}       & \makecell{A ratio of correct GOOSE anomalies that \\the system failed to detect.} & \multicolumn{1}{c|}{21.82\%}        & \multicolumn{1}{c|}{14.55\%}        & \multicolumn{1}{c|}{1.82\%}     & \multicolumn{1}{c|}{21.82\%}        & \multicolumn{1}{c|}{16.36\%}        & \multicolumn{1}{c|}{10.91\%}     & \multicolumn{1}{c|}{25.5\%}        & \multicolumn{1}{c|}{18.18\%}        &   10.9\%   \\ \hline
\textit{Precision} & \makecell{Measures accuracy of detected \\ GOOSE anomalies.} & \multicolumn{1}{c|}{78.18\%}        & \multicolumn{1}{c|}{85.45\%}        & \multicolumn{1}{c|}{98.18\%}     & \multicolumn{1}{c|}{75.43\%}        & \multicolumn{1}{c|}{80.7\%}        & \multicolumn{1}{c|}{85.96\%}     & \multicolumn{1}{c|}{74.5\%}        & \multicolumn{1}{c|}{81.8\%}        &   90.7\%   \\ \hline

%\textit{Accuracy} & \makecell{It measures accuracy of all correctly identified \\ GOOSE messages.} & \multicolumn{1}{c|}{70\%}        & \multicolumn{1}{c|}{80\%}        & \multicolumn{1}{c|}{97.5\%}     & \multicolumn{1}{c|}{67.5\%}        & \multicolumn{1}{c|}{75\%}        & \multicolumn{1}{c|}{82.5\%}     & \multicolumn{1}{c|}{65\%}        & \multicolumn{1}{c|}{75\%}        &   86.25\%   \\ \hline

\textit{F1-Score}  & \makecell{Provides a trade-off between precision\\ and recall.} & \multicolumn{1}{c|}{78.18\%}        & \multicolumn{1}{c|}{85.45\%}        & \multicolumn{1}{c|}{98.18\%}     & \multicolumn{1}{c|}{76.78\%}        & \multicolumn{1}{c|}{82.3\%}        & \multicolumn{1}{c|}{87.5\%}     & \multicolumn{1}{c|}{74.5\%}        & \multicolumn{1}{c|}{81.8\%}        &   90.7\%   \\ \hline
 \cellcolor[HTML]{656565}         & \textbf{\makecell{IEC 61850-based Communication}} & \multicolumn{9}{c|}{\cellcolor[HTML]{FFFFC7}\textbf{SV}}                                                                                                                                                                                               \\ \hline
 \cellcolor[HTML]{656565}     & \makecell{\textbf{LLMs}} & \multicolumn{3}{c|}{ChatGPT 4.0}                                                        & \multicolumn{3}{c|}{Anthropic's Claude 2}                                                        & \multicolumn{3}{c|}{Google Bard/PaLM 2}                                   \\ \hline
      \textbf{\makecell{Metrics}}    & \textbf{\makecell{Description}} & \multicolumn{1}{c|}{without} & \multicolumn{1}{c|}{partial} & \multicolumn{1}{c|}{full} & \multicolumn{1}{c|}{without} & \multicolumn{1}{c|}{partial} & \multicolumn{1}{c|}{full} & \multicolumn{1}{c|}{without} & \multicolumn{1}{c|}{partial} & full \\ \hline
\textit{TPR}       & \makecell{A ratio of correct SV anomalies that were \\correctly identified (also, named recall).} & \multicolumn{1}{c|}{70\%}        & \multicolumn{1}{c|}{95\%}        & \multicolumn{1}{c|}{96.67\%}     & \multicolumn{1}{c|}{50\%}        & \multicolumn{1}{c|}{70\%}        & \multicolumn{1}{c|}{88.3\%}     & \multicolumn{1}{c|}{50\%}        & \multicolumn{1}{c|}{63.3\%}        & 81.6\%     \\ \hline
%\textit{TNR}       & \makecell{A ratio of normal SV data that were \\correctly classified as clean data.} & \multicolumn{1}{c|}{50\%}        & \multicolumn{1}{c|}{85\%}        & \multicolumn{1}{c|}{100\%}     & \multicolumn{1}{c|}{50\%}        & \multicolumn{1}{c|}{80\%}        & \multicolumn{1}{c|}{100\%}     & \multicolumn{1}{c|}{50\%}        & \multicolumn{1}{c|}{60\%}        &  75\%    \\ \hline
\textit{FPR}       & \makecell{A ratio of normal SV data that were \\wrongly identified as anomalies.} & \multicolumn{1}{c|}{50\%}        & \multicolumn{1}{c|}{15\%}        & \multicolumn{1}{c|}{0\%}     & \multicolumn{1}{c|}{50\%}        & \multicolumn{1}{c|}{20\%}        & \multicolumn{1}{c|}{0\%}     & \multicolumn{1}{c|}{50\%}        & \multicolumn{1}{c|}{40\%}        &   25\%   \\ \hline
\textit{FNR}       & \makecell{A ratio of correct SV anomalies that \\the system failed to detect.} & \multicolumn{1}{c|}{30\%}        & \multicolumn{1}{c|}{5\%}        & \multicolumn{1}{c|}{3.33\%}     & \multicolumn{1}{c|}{50\%}        & \multicolumn{1}{c|}{30\%}        & \multicolumn{1}{c|}{11.67\%}     & \multicolumn{1}{c|}{50\%}        & \multicolumn{1}{c|}{36.6\%}        &   18.34\%   \\ \hline
\textit{Precision} & \makecell{Measures accuracy of detected \\SV anomalies.} & \multicolumn{1}{c|}{80.77\%}        & \multicolumn{1}{c|}{95\%}        & \multicolumn{1}{c|}{100\%}     & \multicolumn{1}{c|}{75\%}        & \multicolumn{1}{c|}{91.3\%}        & \multicolumn{1}{c|}{100\%}     & \multicolumn{1}{c|}{75\%}        & \multicolumn{1}{c|}{82.6\%}        &    91.7\%  \\ \hline
%\textit{Accuracy} & \makecell{It measures accuracy of all correctly identified \\SV messages.} & \multicolumn{1}{c|}{65\%}        & \multicolumn{1}{c|}{92.5\%}        & \multicolumn{1}{c|}{97.5\%}     & \multicolumn{1}{c|}{50\%}        & \multicolumn{1}{c|}{72.5\%}        & \multicolumn{1}{c|}{91.25\%}     & \multicolumn{1}{c|}{50\%}        & \multicolumn{1}{c|}{62.5\%}        &   80\%   \\ \hline
\textit{F1-Score}  & \makecell{Provides a trade-off between precision\\ and recall.} & \multicolumn{1}{c|}{75\%}        & \multicolumn{1}{c|}{95\%}        & \multicolumn{1}{c|}{98.3\%}     & \multicolumn{1}{c|}{60\%}        & \multicolumn{1}{c|}{79.2\%}        & \multicolumn{1}{c|}{93.8\%}     & \multicolumn{1}{c|}{60\%}        & \multicolumn{1}{c|}{71.7\%}        &  85.9\%    \\ \hline
\end{tabular}
\end{table*}
\subsection{Case Studies: GOOSE and SV Anomaly Detection}
This section presents the results of the performance evaluation metrics for different LLMs based on the training levels. The formulations of the performance assessment are given in the previous part, including true positive rate (TPR), false positive rate (FPR), false negative rate (FNR), precision, and F1-score metrics. A comparison of anomaly detection results considering the different LLMs (i.e., ChatGPT 4.0, Anthropic's Claude 2, and Google Bard/PaLM 2) with the HITL process is presented in this table. The results show that ChatGPT 4.0 outperforms the two other LLMs in both case studies for anomaly detection as an IDS. A higher TPR indicates a better model, as it is able to identify more of the actual positives. It also shows the detection rate of anomalies, where ChatGPT 4.0 has values of 98.18\% and 96.67\% for detection of anomalies in GOOSE and SV messages, respectively. These percentages are the highest rates in comparison with other LLMs at full training levels. It happened for all other training levels as well. Lower FPR and FNR indicate a superior model, as it is less possible to misclassify positive and negative values, respectively. It occurs for ChatGPT 4.0 considering FPR and FNR in both communications. At full training levels, these values are less than $4\%$ which represents a good performance of this LLM. Also, Claude 2 shows great performance in the detection of normal SV data that was wrongly detected as anomalies. The precision metric represents the accuracy of anomalies detection in GOOSE and SV communications. The precision values for ChatGPT 4.0 and Claude 2 are $100\%$ in comparison with Google Bard ($91.7\%$) in the SV dataset. F1-score is a harmonic mean of precision and recall, which means that it gives equal importance to both the ability of the algorithm to identify true anomalies and its ability to avoid FPs. This metric shows the highest value based on ChatGPT 4.0. The impact of the HITL process can be observed at different training levels in Table~\ref{results}. A portion of the human recommendations are considered for the partial training. Therefore, better performance at different rates, precisions, and F1-scores can be perceived by applying the HITL process. All human recommendations based on the defined attacks/errors are considered at the full training level.

To recap, ChatGPT 4.0 served as the best LLM in comparison with Anthropic's Claude 2 and Google Bard/PaLM 2 in all rates and measurements. However, there are challenges in using LLMs based on the HITL process in cybersecurity studies on digital substations. Cybersecurity anomalies entail a level of complexity that may exceed AI's contextual discernment capabilities. The necessity for LLMs to process sensitive data introduces data privacy and security considerations. AI's enhancement in cybersecurity is hindered by the need for continuous data input, reflecting the dynamic nature of the field. The integrity of anomaly detection in AI is dependent on its training data, with potential inaccuracies manifesting as FPs or FNs. Hence, task-oriented dialogues (ToD) and fine-tuning are posited to enhance anomaly detection accuracy through the provision of structured interactive patterns that augment LLMs' effectiveness in cybersecurity-specific responses. They enable more targeted and context-aware queries, thereby refining the decision-making process. Additionally, they promote an improved feedback mechanism where human experts can iteratively refine AI performance on designated tasks, thereby optimizing its learning trajectory over time.
%~\cite{lee2023knowledge}. 
Rule-based detection systems, known for their effectiveness in identifying known threats, often demonstrate superior results in specific scenarios. However, LLMs bring a unique advantage to the field of anomaly detection. Unlike rule-based systems that rely on predefined criteria, LLMs possess the capability to identify unexpected or novel attacks, a critical feature in the constantly evolving landscape of cybersecurity threats. This ability to detect anomalies that deviate from known patterns or behaviors allows LLMs to address a broader range of potential attacks. Consequently, integrating LLMs into anomaly detection efforts can significantly reduce the manual labor and complexity involved in continuously updating and maintaining rule-based systems, especially in environments where new and unforeseen attack vectors are a constant challenge.
\section{Conclusion}
This paper proposes the use of LLMs based on the HITL process for cybersecurity in substations, as evaluated by various performance metrics. LLMs are employed as IDSs to identify anomalies in communication protocols. An IDS algorithm is converted to text to train datasets for anomaly detection. In comparison, ChatGPT 4.0 outperformed the two other LLMs in all metrics. This LLM demonstrated better precision and performance at different levels of training. These models have privacy issues regarding confidential data. Thus, using the ToD and fine-tuning are necessary to enhance the accuracy of LLMs. In the future, it will be the intention to consider other LLMs with ToD and fine-tuning processes with more attacks and errors to improve the LLMs' efficiency, along with analyses on all multicast messages in digital substations.
\bibliographystyle{IEEEtran}
\bibliography{IEEEabrv,ref}

\end{document}